\documentclass[11pt,nofootinbib,prd,aps,superscriptaddress,preprintnumbers,
  amsmath,amssymb,amsfonts,eqsecnum,secnumarabic,showpacs]{revtex4-1}
\pdfoutput=1
\usepackage{slashed,color,multirow}
\usepackage{graphicx}
\usepackage{subfigure}
\usepackage{dcolumn}
\usepackage{bm}
\usepackage{epsfig}
\usepackage{epstopdf}
\usepackage{setspace}
\usepackage{hyperref}

\newcommand{\beq}{\begin{equation}}
\newcommand{\eeq}{\end{equation}}
\newcommand{\tableref}[1]{Table~\ref{tab:#1}}

\newcommand{\figref}[1]{Fig.~\ref{fig:#1}}
\newcommand{\beqa}{\begin{eqnarray}}
\newcommand{\eeqa}{\end{eqnarray}}

\def\bsp#1\esp{\begin{split}#1\end{split}}

\begin{document}

\def\UCI{Department  of  Physics  and  Astronomy,
  University  of  California,  Irvine,  CA  92697,  USA}
\def\Weiz{Department of Particle Physics and Astrophysics,
   Weizmann Institute of Science, Rehovot 76100, Israel}
\def\KITa{Institut fur Theoretische Teilchenphysik,
  Karlsruhe Institute of Technology, Engesserstra\ss e 7,
  D-76128 Karlsruhe, Germany}
\def\KITb{Institut fur Kernphysik, Karlsruhe Institute of Technology,
  Hermann-von-Helmholtz-Platz 1,
  D-76344 Eggenstein-Leopoldshafen, Germany}
\def\ParisA{Sorbonne Universit\'es, UPMC Univ.~Paris 06,
  UMR 7589, LPTHE, F-75005, Paris, France}
\def\ParisB{CNRS, UMR 7589, LPTHE, F-75005, Paris, France}

\title{Gluino Meets Flavored Naturalness }
\author{Monika \surname{Blanke}}
\affiliation{\KITb}
\affiliation{\KITa}
\author{Benjamin~\surname{Fuks}}
\affiliation{\ParisA}
\affiliation{\ParisB}
\author{Iftah \surname{Galon}}
\affiliation{\UCI}
\author{Gilad \surname{Perez}}
\affiliation{\Weiz}

\date{\today}

\preprint{TTP15-043}
\preprint{UCI-TR/2015-22}

\begin{abstract}
We study constraints from LHC run I on squark and gluino masses in the presence
of squark flavor violation. Inspired by the concept of `flavored naturalness',
we focus on the impact of a non-zero stop-scharm mixing and mass splitting in the
right-handed sector. To this end, we recast four searches of the ATLAS and CMS
collaborations, dedicated either to third generation squarks, to gluino and
squarks of the first two generations, or to charm-squarks. In the absence of
extra structure, the mass of the gluino provides an additional source of fine
tuning and is therefore important to consider within models of flavored
naturalness that allow for relatively light squark states. When combining the
searches, the resulting constraints in the plane of the lightest squark and
gluino masses are rather stable with respect to the presence of flavor-violation, and do
not allow for gluino masses of less than~1.2 TeV and squarks lighter than about
550~GeV. While these constraints are stringent, interesting models with sizable stop-scharm mixing and a relatively light squark state are still viable and
could be observed in the near future.
\end{abstract}

\maketitle

\section{Introduction}

While the Large Hadron Collider (LHC) has just begun its second period of data
taking, its first run has been an experimental success with many new
measurements at an energy regime unexplored beforehand.
In particular, the search for new phenomena has been given a boost with the
discovery of a new particle, the celebrated Higgs boson~\cite{Aad:2012tfa,
Chatrchyan:2012xdj}.
Theoretically, however, we are still in the dark, as most
searches performed at the LHC
seem to be consistent with the Standard Model (SM) predictions.
This includes the measurements of the Higgs couplings~\cite{Hcouplings},
the searches for new physics at the energy frontier
with the ATLAS and CMS experiments and at the luminosity frontier
with the LHCb experiment\footnote{
Although the recent discovery of pentaquark-like states~\cite{Aaij:2015tga}
is explained within the SM, the recent LHCb flavor anomalies might imply new
physics~\cite{Aaij:2014pli,Aaij:2014ora,Aaij:2015yra}.}.
In the absence of new physics signals, the naturalness 
argument that
motivates the possible observation of new dynamics at the TeV scale seems
slightly less appealing as a guiding principle 
(see Ref.~\cite{Giudice:2013nak} for a general status review
and Ref.~\cite{Feng:2013pwa} for a focus on supersymmetry).
One of the main tasks of the next high-energy LHC runs at 13~TeV and 14~TeV will
hence be to shed more light on the electroweak symmetry breaking mechanism and
to estimate to what extent the Higgs-boson mass is fine-tuned.

One of the robust features of all natural extensions of the SM is the presence of
top partners. These act to screen away the quadratic sensitivity of the
Higgs-boson mass to the ultraviolet (UV) scales due mostly to the large top Yukawa coupling.
Naively, one might expect that flavor physics and naturalness are 
two decoupled concepts.
However, even within a minimal top partner sector, 
the definition of the flavor
structure of the model could be non-trivial.
The mass-eigenstates of the theory could be non-pure top partners and still
yield a sufficient cancellation of the UV-sensitive quantum contributions to
the Higgs-boson mass. 
In this way, even a model exhibiting a
single top-partner might incorporate large flavor- and CP-violating effects.
This possibility, however, is typically ignored, due to
prejudices and a possibly too simplistic interpretation of the bounds stemming
from low-energy flavor-changing neutral current processes.
Indeed, most studies on naturalness have assumed 
either flavor universality among the partners
or an approximate $U(2)$ symmetry which acts 
on the partners of the first two
generations.
Nonetheless, 
a thorough analysis of the constraints arising from $D-\bar D$ and
$K-\bar K$ mixing has shown that the degeneracy of the partners is not required
for models of down alignment~\cite{Gedalia:2012pi}, and
such frameworks in which new physics couplings are non-diagonal in flavor-space
have been considered both in the context of
supersymmetry~\cite{Nir:1993mx, Galon:2013jba} and Higgs
compositeness~\cite{Delaunay:2010dw, Azatov:2014lha}.

Taking supersymmetry as an illustrative example, the non-degeneracy of the
partners is even more appealing as the direct experimental bounds on second
generation squarks are rather weak, their masses being only constrained to be
larger than 500~GeV.
This is a consequence of the underlying ingredients of all
supersymmetry searches which are mainly sensitive either to `valence'
squarks or to third generation squarks~\cite{Mahbubani:2012qq}.
If the supersymmetric top-partners are not flavor-eigenstates
but rather admixtures of stops and scharms,
then the signatures of supersymmetric events could change dramatically.
In particular, the typically sought signatures such as top-quark pairs and
missing transverse energy ($\slashed E_T$)
could be exchanged for charm-jet pairs
and top-charm pairs plus $\slashed E_T$.
This has led to the concept of supersymmetric `flavored naturalness'. 
{Despite the non-trivial flavor structure} of the top sector, 
the level of fine tuning of these setups is similar to 
more conventional supersymmetric scenarios with
pure-stop mass eigenstates and sometimes even improved~\cite{Blanke:2013uia}. 
In addition, it has been shown that low-energy electroweak and 
flavor physics still allow for large deviations from a minimal flavor structure in the
squark sector~\cite{Heinemeyer:2004by,Cao:2006xb,Bozzi:2007me,Buchalla:2008jp,
Fuks:2008ab, Hurth:2009ke,Fuks:2011dg,Agrawal:2013kha,Arana-Catania:2014ooa,Kowalska:2014opa,Backovic:2015rwa,
DeCausmaecker:2015yca}.

Complementary, the gluino state included in any supersymmetric extension of the
SM leads to an independent source of fine tuning~\cite{Brust:2011tb}.
This result is a combination of the rather strong bounds on the gluino mass, 
and the genuine gluino loop-diagram contributions to all scalar masses 
(and in particular to the squark masses $m_{\tilde q}$).
Indeed, the gluino mass $m_{\tilde g}$ which is constrained to be larger than about
1~TeV, and even more in some specific setups, 
implies a naturalness relation between the 
squark and gluino masses,
\beq
  m_{\tilde g}\lesssim 2 {m_{\tilde q} } \ .
\eeq
In this paper we focus on gluino phenomenology and study how the presence of a
not that heavy gluino in the theory can give rise to bounds on flavored
naturalness, and on non-minimal flavor mixing in the squark sector.
In the case of squark flavor-violation, 
several new supersymmetric decay channels become relevant and new
signatures could be expected.
Although there is no specific experimental search dedicated to this
non-minimally flavor-violating supersymmetric setup,
the panel of signatures that can arise is large enough such that we can
expect to be able to derive constraints from standard
analyses that have been designed from flavor-conserving supersymmetric
considerations, as already depicted in previous prospective
studies~\cite{Han:2003qe,Bozzi:2007me,Hiller:2008wp,Fuks:2008ab,Kribs:2009zy,
Bartl:2009au,Hurth:2009ke,Bartl:2010du,Bruhnke:2010rh,Bartl:2011wq,Bartl:2012tx,
Fuks:2011dg,Blanke:2013uia,DeCausmaecker:2015yca}.
We therefore investigate how standard searches
for squarks and gluinos can be sensitive to a non-trivial flavor structure in
the squark sector.
In particular, we focus on four searches of the
ATLAS and CMS collaborations, dedicated either to third generation
squarks~\cite{Chatrchyan:2013xna}, to gluino and squarks from the
first two generations~\cite{Aad:2013wta,Chatrchyan:2014lfa}, 
or to charm-squarks~\cite{Aad:2015gna}.
This is only a representative subset of the vast experimental wealth 
of searches, but it is sufficient to derive meaningful results
that constrain flavored naturalness
since it covers 
all topologies that can arise in our non-minimally flavor-violating
setup.

This work is structured as follows:
The simplified model description that has been used throughout this analysis is
introduced in Section~\ref{sec:SimplifiedModels}.
Section~\ref{sec:recast} describes the reinterpretation procedure including
the simulation setup and a concise summary of the experimental searches under
consideration.
The results are discussed in Section~\ref{sec:results}, while our conclusions are
presented in Section~\ref{sec:conclusions}. 
Details regarding the implementation of the ATLAS-SUSY-2013-04 search in the
reinterpretation framework that we have used are provided in
Appendix~\ref{sec:validation}, while those related to the other considered
searches can be found in Ref.~\cite{Dumont:2014tja}

\section{Theoretical Framework: a Simplified Model for Studying
  Gluino Flavor Violation}
\label{sec:SimplifiedModels}
Following a simplified model approach such as those traditionally employed by
the ATLAS and CMS collaborations~\cite{Alwall:2008ag,Alves:2011wf}, we consider
a supersymmetric extension of the SM where only a subset of the superpartners
feature masses accessible at the LHC.
Effectively, we supplement the SM by a neutralino state
$\tilde \chi^0$ which is the Lightest Supersymmetric Particle (LSP),
one gluino state $\tilde g$ and two up-type squark states
$\tilde u_1$ and $\tilde u_2$.
The latter are mass eigenstates which are linear combinations 
of the right-handed stop and scharm
flavor-eigenstates,
\renewcommand{\arraystretch}{1.2} \beq
  \begin{pmatrix} \tilde u_1 \\ \tilde u_2 \end{pmatrix} =
  \begin{pmatrix}
    \phantom{-}\cos\theta^R_{23} & \sin\theta^R_{23} \\
    -\sin\theta^R_{23} & \cos\theta^R_{23}
  \end{pmatrix}
  \begin{pmatrix} \tilde c_R \\ \tilde t_R \end{pmatrix}
   \qquad\text{with}\qquad 0 \le \theta^R_{23} \le \pi/4 \ .
\eeq\renewcommand{\arraystretch}{1.0}%
Note that the squark mixing angle lies in the $[0,\pi/4]$ range so that the $\tilde u_1$ ($\tilde u_2$) state always contains a dominant scharm (stop) component.
Throughout our analysis, we consider an extremely light neutralino 
with a mass fixed at $m_{\tilde \chi^0} = 1$~GeV. 
This assumption maximizes the
amount of missing energy to be produced in signal events, and thus
represents a favored scenario for any analysis relying on large missing energy signatures. 
Consequently, the results of this work represent a conservative estimate of the
LHC sensitivity to the studied setup.

We define our model parameter-space by the three remaining masses, namely the
squark and gluino
masses $m_{\tilde u_1}$, $m_{\tilde u_2}$ and $m_{\tilde g}$, and the squark
mixing angle $\theta^R_{23}$. 
Although strong bounds on the parameter-space
could be derived from flavor physics observables even with just two `active'
squark states, all flavor constraints are, in practice, only sensitive to the
quantity
\beq
\left(\delta^u_{23}\right)_{RR} \approx
  \left|\frac{m_{\tilde u_1} - m_{\tilde u_2}}
    {m_{\tilde u_1} + m_{\tilde u_2}}\right|^2\ 
  \sin\left(2\theta^R_{23}\right)\ ,
\eeq
in the mass insertion approximation~\cite{Raz:2002zx,Nir:2002ah} that is suitable
for relatively small mass splittings and mixings.
Since flavor bounds on the squark masses and mixings in the right-right sector are
mild {and arise mostly from $D$ physics~\cite{Ciuchini:2007ha, Isidori:2010kg}, flavor violation involving the right-handed stop and scharm is essentially unconstrained by flavor data, as long as the mixing between the stop and the up squark is assumed to be small. Therefore} the squark mass difference
$\Delta m =m_{\tilde u_2}-m_{\tilde u_1}$ and the sine of the mixing angle
$\sin\theta^R_{23}$ could both be large, with related implications on squark
pair-production cross sections~\cite{Bozzi:2007me}. Furthermore,
LHC searches are potentially sensitive to the $\Delta m$ and
$\sin\theta^R_{23}$ parameters separately, in contrast to flavor
constraints which cannot disentangle the two~\cite{Calibbi:2015kja}\footnote{
This was pointed out in Ref.~\cite{Calibbi:2015kja} in the context of simplified models featuring a flavorful slepton sector. 
In practice, due to the highly restrictive lepton flavor bounds, 
their analysis included scenarios where only one parameter
($\Delta m$ or $\sin\theta_{23}^R$) was large enough to induce
an observable effect in the slepton searches of the first LHC run.
}.

The parameter-space can be divided into two domains depending on whether $\tilde u_1$ or $\tilde u_2$ is the lightest squark.
For the sake of a clearer discussion and in order to put
these two setups on an equal footing, we define two series of scenarios, both
parameterized by a $\big(m_{\tilde g}, m_{\tilde u}, \Delta m,
\theta_{23}^R\big)$ tuple, and distinguished by the sign of $\Delta m$.
For the first series of scenarios (that we denote by {\bf S.I}), the
$m_{\tilde u}$ parameter is identified with the
$\tilde u_1$ mass, while for the second series of scenarios (that we denote by
{\bf S.II}), the two squark masses are interchanged and the stop-dominated
$\tilde u_2$ squark is now the lightest squark state with a mass given by
$m_{\tilde u}$.
This can be summarized, for given values of $\theta^R_{23}$ and $m_{\tilde g}$, as
\beq
\begin{cases}
\text{Scenarios of type {\bf S.I}}\\
m_{\tilde u} = m_{\tilde u_1} < m_{\tilde u_2} \\
\Delta m = m_{\tilde u_2} - m_{\tilde u_1} >0 \\
\end{cases}
\qquad\text{and}\qquad
\begin{cases}
\text{Scenarios of type {\bf S.II}}\\
m_{\tilde u} = m_{\tilde u_2} < m_{\tilde u_1} \\
\Delta m = m_{\tilde u_2} - m_{\tilde u_1} <0 \\
\end{cases} \ .
\eeq

In order to study the gluino effects on the constraints that can be imposed on
flavored naturalness models, we perform a scan
of the above parameter-space. The range in which each physical parameter of the
model description is allowed to vary is given in \tableref{scan_params}.
The $\Delta m = 0$ case deserves a clarification.
If the two squarks are indeed entirely degenerate, then the mixing can be
rotated away as a consequence of the $U(2)$ symmetry of the two squark
mass-squared matrix. In this case, the mixing has thus no physical meaning. By
imposing $\Delta m = 0$, we in fact mean
that the splitting between the squark states does not manifest itself in LHC
processes and is still larger than the width of the squarks so that oscillation and interference effects
are unimportant.

\renewcommand{\arraystretch}{1.2}
\begin{table}
\renewcommand{\tabcolsep}{0.30cm}
\begin{tabular}{c||c|c|c}
Parameter & Minimum value & Maximum value & Step \\
\hline \hline
$m_{\tilde \chi}$ [GeV] &    1 &     1          & 0   \\ \hline
$m_{\tilde g}$ [GeV]    &  800 & 2000           & 100 \\ \hline
$m_{\tilde u}$ [GeV]    &  400 & $m_{\tilde g}$ & 100 \\ \hline
$\Delta m$ [GeV]        &  -500   & 500            & 100 \\ \hline
$\theta^R_{23}$         &  0   & $\pi/4$        & $\pi/20$\\
\end{tabular}
\caption{Summary of the parameter-space of our simplified model. We present each
physical parameter together with details associated with the scan that we have
performed.}
\label{tab:scan_params}
\end{table}
\renewcommand{\arraystretch}{1.0}

\section{Monte Carlo Simulations and LHC Analysis Reinterpretation Details}
\label{sec:recast}

\subsection{Technical setup and general considerations}
\label{sec:generalities}
To determine the LHC sensitivity to the class of models introduced in
Section~\ref{sec:SimplifiedModels}, we reinterpret the results of several ATLAS
and CMS searches for supersymmetry for each point of the parameter-space scan
defined in Table~\ref{tab:scan_params}. 
Technically, we have started by implementing
the simplified model described above into {\sc FeynRules}~\cite{Alloul:2013bka},
exported the model information in the UFO format~\cite{Degrande:2011ua} and
have then made use of the {\sc MadGraph5}\_aMC@NLO~\cite{MG5} framework for event generation. 
The description of the QCD environment (parton showering and
hadronization) has been achieved with the use of the {\sc Pythia 6}~\cite{PYTHIA6}
package. Next, we apply a detector response emulator to the simulation results
by means of the {\sc Delphes}~3 program~\cite{Delphes}, that internally relies on
the anti-$k_T$ jet algorithm~\cite{AntiKt} as implemented in the
{\sc FastJet} software~\cite{Cacciari:2011ma} for object reconstruction. For
each of the recasted analyses, the {\sc Delphes} configuration has been
consistently tuned to match the setup depicted in the experimental
documentation~\cite{Dumont:2014tja}.
Finally, we have used the {\sc MadAnalysis}~5 framework~\cite{Conte:2012fm,Conte:2014zja}
to calculate the signal efficiencies for the different search strategies, and
to derive 95\% confidence level (CL) exclusions with the CLs method~\cite{CLs}.

As most of the LHC analyses under consideration rely on a proper description of
the jet properties, we have merged event samples containing up to one extra jet
compared to the Born process and accounted for the possible double-counting arising
from radiation that could be described both at the level of the matrix element and
at the level of the parton showering by means of the Mangano (MLM)
scheme~\cite{Mangano:2006rw,Alwall:2008qv}. 
Moreover, we have normalized the cross-section for the signal samples to
the next-to-leading-order (NLO) accuracy in QCD.
Nevertheless, as flavor effects on supersymmetric production cross-sections at NLO
have yet to be calculated, we have taken a very conservative approach and
applied a global $K$-factor of 1.25 to all leading-order results
that have been obtained with {\sc MadGraph5}\_aMC@NLO.

In the above simulation chain, we employ a detector simulation 
which is much simpler than those 
of the CMS and ATLAS experiments.
It cannot therefore genuinely account for the
full complexity of the real detectors. 
For instance, event cleaning requirements or
basic object quality criteria cannot be implemented in {\sc Delphes}. 
While those are expected to only have a small
impact on the limits that are derived, 
it is important to bear in mind that
related uncertainties exist.
Furthermore, the searches we are focusing on are multichannel searches, and the
experimental limits are often derived after combining all channels. 
The statistical models that are used in the official exclusions are however not publicly
available, so that we have made the approximation of {considering each search
channel independently} and computed our limits by restricting ourselves to the
channel yielding the strongest exclusion. 
In this way, we have omitted all
correlations that could improve the bounds.
Finally, in some of the considered searches, 
the background estimation in the various
signal regions depends on an extrapolation from designated control regions.
Consequently, the possibility of control region contamination by signal events
should also be taken into account. 
This contamination, however, depends on the signal model
being explored, and the information on how we should quantify it is not public.
This has therefore not been pursued in our work.

The combined effect of all these features leads to results that are 
compatible with the experimental ones within
$10\%-20\%$~\cite{Dumont:2014tja}. {We stress however that the {\it relative} uncertainty is much smaller. Many of the errors arising from our simplified recast should lead to an overall mis-estimation of a given bound. Yet we do not expect these recast errors to be sensitive to the size of the flavor mixing. In other words, when taking ratios of bounds (which we effectively do when considering how bounds change), this systematic uncertainty should largely drop out.}

We have verified the consistency of our methodology for the analyses under
consideration, and validated in this way our reimplementation procedure. 
More details are given in the rest of this section, in
Appendix~\ref{sec:validation} as well as in Ref.~\cite{Dumont:2014tja}.

\subsection{\boldmath ATLAS: multijets + $\slashed{E}_T$ + lepton veto}

The ATLAS-SUSY-2013-04 search~\cite{Aad:2013wta} is a supersymmetry-oriented
search which focuses on a multijet signature accompanied by large missing
transverse energy and no isolated hard leptons (electron or muon).
In the context of our model description of Section~\ref{sec:SimplifiedModels},
it targets the production of gluino pairs, squark pairs or gluino-squark associated
pairs that subsequently decay into missing transverse energy and jets.
The selection strategy relies on dedicated multijet triggers with a minimal
requirement of five (six) very energetic central jets with a transverse
energy $E_T > 55$~GeV ($E_T > 45$~GeV) and a pseudorapidity satisfying
$|\eta|<3.2$. Events are then collected into two types of signal regions, the
so-called `multijet + flavor stream' and `multijet + $M^\Sigma_J$ stream'
categories, which yields a total of 19 overlapping signal regions.

In the `multijet + flavor stream' signal region category, the events are classified
according to requirements on the number of jets exhibiting specific properties.
In a first set of regions, the events are required to feature exactly 8 (8j50),
9 (9j50) or at least 10 ($\ge$10j50) jets with a transverse momentum
$p_T > 50$~GeV and a pseudorapidity $|\eta|<2$. In a second set of regions,
they are constrained to contain exactly 7 (7j80) or
at least 8 ($\ge$8j80) jets with a transverse momentum $p_T > 80$~GeV and a
pseudorapidity $|\eta|<2$. Except for the $\ge$10j50 region, a further
subdivision is made according to the number of $b$-tagged jets (0, 1 or at
least 2) with a pseudorapidity $|\eta|<2.5$ and a transverse momentum
$p_T >40$~GeV.
The signal selection strategy finally relies on the missing
transverse energy significance, $\slashed{E}_T/\sqrt{H_T} > 4~\rm{GeV}^{1/2}$,
where $H_T$ is defined as the sum of the hadronic transverse energy of all jets
with $E_T$ larger than $50$~GeV. 
For SM processes, this variable is expected to be small,
and almost insensitive to the jet multiplicity. 
A further background reduction is thus obtained by requiring
the same, relatively high missing transverse-energy significance
in all signal regions.

The `multijet + $M^\Sigma_J$ stream' signal region category relies on an extra
variable, $M^\Sigma_J$, that is defined as the invariant mass obtained after
combining the momenta of all fat jets (whose radius is $R=1$) with a
transverse momentum larger than 100~GeV and a pseudorapidity smaller than 1.5 in
absolute value.
Unfortunately, the {\sc MadAnalysis}~5 framework is currently unable
to handle fat jets. Consequently, we refrain ourselves from implementing this
type of signal region flow in our recasting procedure.

This work features the first use of a reimplementation of the ATLAS-SUSY-2013-04
search in the {\sc MadAnalysis}~5 framework. Details on its validation are
therefore given in Appendix~\ref{sec:validation}.

\subsection{\boldmath CMS: single lepton + at least four jets (including at least
one $b$-jet) + $\slashed{E}_T$}

The CMS-SUS-13-011 search~\cite{Chatrchyan:2013xna} is a stop search that targets
stop pair-production and two possible decay modes of the stop,
$\tilde t \to t\tilde\chi^0 \to W^+b\tilde\chi^0$ and
$\tilde t \to b\tilde\chi^+ \to bW^+\tilde\chi^0$, that lead to similar
final-state topologies. 
In the model description of
Section~\ref{sec:SimplifiedModels} we have assumed that the charginos are
decoupled, which is a fair assumption if the $\mu$-term
is large.
Nevertheless, we still include this search in our analysis as
its signal regions cover signatures
which are related  to the top-neutralino decay of the stop.
The CMS-SUS-13-011 search targets events
where one of the $W$-bosons decays hadronically, while the other one decays leptonically, into an electron or a muon (the $\tau$ channel is ignored).
It contains two analysis flows, a first one using a predefined selection strategy
and that is hence `cut-based', and a second one relying on a boosted decision
tree (BDT) technique. Although the BDT analysis provides a sensitivity that is
40\% better, the absence of related public information prevents the community
from making use of it for phenomenological purposes. We therefore focus only on
the cut-based analysis strategy.

The object definition and event preselection criteria require the presence of a
single isolated lepton with a transverse momentum $p_T > 30$~GeV (25~GeV) and a
pseudorapidity $|\eta|<1.4$~(2.1) in the case of an electron (a muon).
Moreover,
no jet can be found in a cone of $R=0.4$ centered on the lepton. A veto is
further enforced on events featuring an additional (loosely) isolated lepton or
a track with an electric charge opposite to the one of the primary
lepton, as well as on events containing hadronic taus. Furthermore, at least
four jets with a transverse momentum $p_T > 30$~GeV and a pseudorapidity
$|\eta|< 2.4$ are required, with at least one of them being
$b$-tagged. The preselection finally imposes that $\slashed{E}_T>100$~GeV, that
the azimuthal angle between the missing momentum and the first two leading jets
is above 0.8 and that the transverse mass constructed from the lepton and the
missing momentum is larger than 120~GeV.

Various signal regions are then defined from several considerations. First, one
designs categories dedicated to probe each of the two considered
stop decay modes, $\tilde t\to t\tilde\chi^0$ and $\tilde t\to b\tilde\chi^+$.
To this aim, one imposes constraints on the hadronic top reconstruction quality
in the case of the region category related to the top-neutralino stop decay
as well as on the amount of missing energy. In the
case of a stop into a neutralino decay, four overlapping signal regions are defined
after requiring $\slashed{E}_T$ to be larger than 150, 200, 250 and 300~GeV,
respectively. In the case of a stop into chargino decay, four regions are again
defined, but using missing energy thresholds of 100, 150, 200 and 250~GeV.
Next, the categories are further subdivided into regions whose goal is to probe
large or small mass differences $\Delta M$ between the stop and the LSP.
Large $\Delta M$ regions are defined by enforcing the transverse variable
$M_{T2}^W$~\cite{Bai:2012gs} to be larger than 200~GeV and the leading $b$-jet
to have a $p_T$ greater than 100~GeV, this last criterion being only relevant
for the case of a stop decay into a chargino.

Information on the implementation and validation of this analysis in the
{\sc MadAnalysis}~5 framework can be found in Ref.~\cite{Dumont:2014tja}
and on {\sc Inspire}~\cite{inspire11}.

\subsection{\boldmath CMS: at least 3~jets + $\slashed{E}_T$ + lepton veto}
The CMS-SUS-13-012 analysis~\cite{Chatrchyan:2014lfa} is a search for supersymmetry, which focuses on the pair-production of gluinos and squarks. 
The main final state sought in this search is comprised of a multijet system and missing
transverse energy, without any isolated leptons. It is hence directly sensitive
to our simplified models in which such signatures would be copiously produced.

The event selection requires at least three jets with a transverse momentum
$p_T>50$~GeV and a pseudorapidity satisfying $|\eta|<2.5$. The total hadronic
activity of the events is then estimated by means of the $H_T$ variable defined
as the scalar sum of the transverse momenta of all jets satisfying the above
requirements. The amount of missing transverse energy in the events is computed
via the $\vec{\slashed{H}}_T$ vector obtained by a vector sum of the transverse
momenta of all jets with $p_T>30$~GeV and a pseudorapidity smaller than 5 in
absolute value. The analysis requires that $H_T > 500$~GeV,
$\slashed H_T > 200$~GeV (where $\slashed H_T =|\vec{\slashed  H}_T|$) and events
in which one of the three hardest jets is aligned with $\vec{\slashed{H}}_T$ are
vetoed by requiring $|\Delta\phi(p_T^j,~\vec{\slashed  H}_T)| > 0.5$ for the two
hardest jets, and $|\Delta\phi(p_T^j,~\vec{\slashed  H}_T)| > 0.3$ for the third hardest one. The selection strategy finally includes a veto on any isolated
lepton whose transverse momentum is larger than 10~GeV.
The events that pass these selection criteria are then categorized into 36
non-overlapping signal regions defined by the number of jets and the value of
the $H_T$ and $\slashed H_T$ variables.

Information on the implementation and validation of this analysis in the
{\sc MadAnalysis}~5 framework can be found in Ref.~\cite{Dumont:2014tja} and
on {\sc Inspire}~\cite{inspire12}.

\subsection{ATLAS: scharm pair-production using charm-tagging + lepton veto}
\label{subsec:scharm_pair_search}
The ATLAS-SUSY-2014-03 search~\cite{Aad:2015gna} is a supersymmetry
search that looks for scharm pair production followed by the $\tilde c\to c
\tilde \chi^0$ decay. The final state is thus comprised of two charm-jets and
missing energy. The experimental analysis therefore targets events that
are required to present a large amount of missing transverse energy,
$\slashed{E}_T > 150$~GeV, at least two very hard jets with transverse
momenta greater than 130 and 100~GeV, respectively, and no isolated leptons.
The two jets are then
demanded to be $c$-tagged. This constitutes the main novel feature of this
search, that involves dedicated charm-tagging techniques based on algorithms
optimized with neural networks. Additional requirements are finally applied
by constraining the so-called contransverse mass~\cite{Tovey:2008ui}.

Mimicking charm-tagging algorithms is beyond the ability of our simplified
detector emulation that relies on {\sc Delphes}. We therefore recast this search
following a different strategy, not based on the {\sc MadAnalysis}~5 framework.
A very conservative (over)-estimate of the bounds is instead derived from the
cross-section limits presented in the experimental
publication~\cite{Aad:2015gna} that we compare to theoretical predictions for
the production cross section of a system made of two charm quarks and two
neutralinos that originate from the decay of two superpartners. More precisely,
the theoretical cross section is calculated from the sum of the
production cross sections $\sigma_{xy}$ of any pair of superpartners $x$
and $y$, the individual channels being reweighted by the corresponding branching
ratios (BR) so that a final state made of two charm quarks and two neutralinos
is ensured,
{\small \begin{eqnarray}
 \sigma^{\rm SUSY}(2c + \slashed{E}_T) & = &
  \sum_{i,j=1,2}\bigg[
     \Big(\frac{\sigma_{\tilde u_i\tilde u_j^*}+\sigma_{\tilde u_j\tilde u_i^*}}
        {1+\delta_{ij}}+
      \sigma_{\tilde u_i\tilde u_j}+ \sigma_{\tilde u_i^*\tilde u_j^*}\Big)\times
    {\rm BR}_{\tilde u_i\to c\tilde\chi_1^0} \times
    {\rm BR}_{\tilde u_j\to c\tilde\chi_1^0} \bigg]\\
&&+ \ \sigma_{\tilde g\tilde\chi_1^0} \times \sum_{i=1,2} \Big[
     {\rm BR}_{\tilde g\to c\tilde u_i}\times
     {\rm BR}_{\tilde u_i\to c\tilde\chi_1^0} \Big]\nonumber\\
&  &+\  \sum_{i,j=1,2}
    \Big(\sigma_{\tilde g\tilde u_j}\!+\!\sigma_{\tilde g\tilde u_j^*}\Big)
     \!\times\!
     \Big[
     {\rm BR}_{\tilde g\to c\tilde u_i}\!\times\!
     {\rm BR}_{\tilde u_i\to c\tilde\chi_1^0}\nonumber \\
&& \hspace*{3.8cm}
    + \Big({\rm BR}_{\tilde g\to c\tilde u_i} \!\times\!
     {\rm BR}_{\tilde u_i\to t\tilde\chi_1^0} \!+\!
       {\rm BR}_{\tilde g\to t\tilde u_i} \!\times\!
       {\rm BR}_{\tilde u_i\to c\tilde\chi_1^0}\Big) \!\times\!
      {\rm BR}_{\tilde u_j\to c\tilde\chi_1^0}\Big]\nonumber \\
 &&  +\ \sigma_{\tilde g\tilde g} \!\times\!
   \sum_{i,j=1,2} \Big[{\rm BR}_{\tilde g\to c\tilde u_i}\!\times\!
         {\rm BR}_{\tilde g\to c\tilde u_j} +
    {\rm BR}_{\tilde g\to t\tilde u_i} \!\times\!
    {\rm BR}_{\tilde u_i\to c\tilde\chi_1^0} \!\times\!
    {\rm BR}_{\tilde g\to t\tilde u_j} \!\times\!
    {\rm BR}_{\tilde u_j\to c\tilde\chi_1^0} \nonumber\\
&&
  \hspace*{2.5cm}+\ 2 {\rm BR}_{\tilde g\to t\tilde u_i} \!\times\!
      {\rm BR}_{\tilde g\to c\tilde u_j} \Big(
     {\rm BR}_{\tilde u_i\to c\tilde\chi_1^0} + 
 {\rm BR}_{\tilde u_i\to t\tilde\chi_1^0} \!\times\!
    {\rm BR}_{\tilde u_j\to c\tilde\chi_1^0}\Big)
\Big]\ . \nonumber
\end{eqnarray}}%
We then implicitly (and
incorrectly) assume that all the other requirements of the ATLAS analysis
described above are fulfilled. In particular, the fact that the neutralino is
almost massless in our parameterization enforces a large amount of missing
energy.

\section{Results}\label{sec:results}

In this section, we present and discuss the main results of our reinterpretation
study. As a preliminary, we introduce two concepts that we
call `{\it signal migration}' and `{\it signal depletion}'. Each of the recasted
experimental analyses targets several event topologies which are assigned to one
(exclusive) or more (inclusive) signal regions. These signal regions then serve
as exclusion channels for the new physics scenarios that are probed. 
In our case, the typical effect of the flavor mixing and the squark mass splitting 
will be to modify the branching fractions of the particles,
and perhaps also the rate of some production processes. 
Consequently, signal regions
that are usually largely populated by signal events in the case
where there is no squark mixing can turn out to be depleted from events once
flavor violation in the squark sector is allowed, and conversely, signal regions
that are not sensitive to any supersymmetric signal in the flavor-conserving
case can become populated. 
One thus expects a migration of the signal across
the different regions with squark flavor violation.

As a concrete example, we compare the event topologies resulting from
the decay $\tilde t\to q \tilde\chi^0$ in a flavor-conserving model
to the case of a model with stop-scharm mixing.
If the stop decays solely into tops, one expects signatures which
would include a $b$-tagged jet, in addition to either two other jets or 
a lepton-neutrino pair that originates from the decay of the $W$-boson.
Alternatively, if the stop can also decay into charm-jets,
the jet multiplicity distribution of the stop decay products peaks to a smaller
value.
Furthermore, a decay into a charm jet, rather than into a top quark,
is not bounded by the top mass and can result in larger neutralino energies
which manifest as a signature with a larger amount of missing energy.

The direct interpretation of this example in the context of the searches
under consideration is simple. 
In comparing a flavor-conserving to a flavor-violating scenario, one expects
that signal regions which are defined according to
requirements on the missing energy, the number of isolated leptons,
the number of jets and the number of $b$-tagged jets,
to redistribute signal events between one another.
This migration of signal events could cause a signal region depletion or population.

In the rest of this section, we present and discuss our results
in the $m_{\tilde u}-m_{\tilde g}$ plane for the model parameterization
described in Section~\ref{sec:SimplifiedModels}.

\subsection{\boldmath ATLAS: multijets + $\slashed{E}_T$ + lepton veto}

\begin{figure}
\centering
\includegraphics[width=0.45\textwidth]{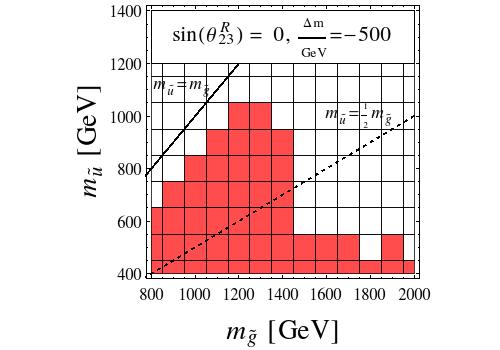}\\[1.5cm]
\includegraphics[width=0.45\textwidth]{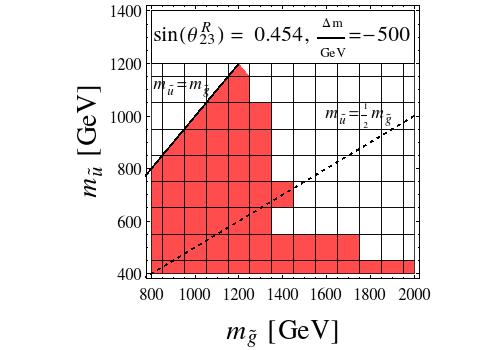}
\includegraphics[width=0.45\textwidth]{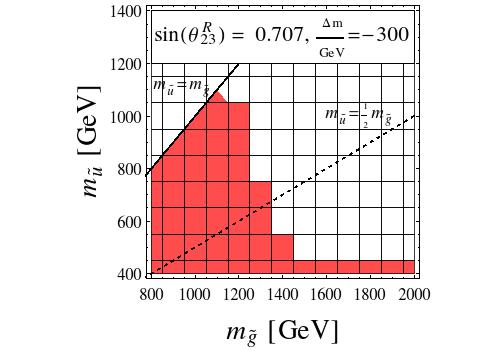}\\[1.5cm]
\includegraphics[width=0.45\textwidth]{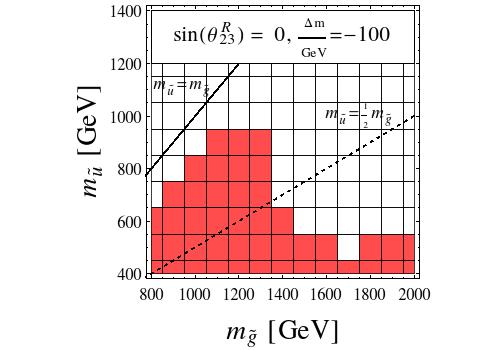}
\includegraphics[width=0.45\textwidth]{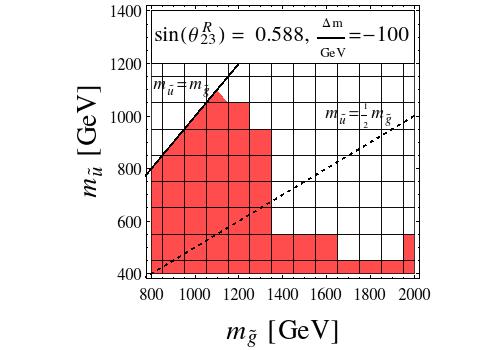}
\caption{
  Sensitivity of the ATLAS multijet + $\slashed{E}_T$ + lepton veto search
  of Ref.~\cite{Aad:2013wta} for different values of the $\Delta m$ and
  $\theta^R_{23}$ parameters (the exact values being indicated in the top bar of
  each subfigure). The excluded regions are shown in red in the $(m_{\tilde u},
  m_{\tilde g})$ plane, where $m_{\tilde u}$ is the mass of the lightest squark
  (a stop-like squark here since $\Delta m < 0$).
  The upper panel describes the reference scenario in which
  the lightest squark is a pure stop state and the scharm is almost decoupled.
}
\label{fig:results_ATLASmultijet_del_m}
\end{figure}

The ATLAS-SUSY-2013-04 search~\cite{Aad:2013wta} has been designed to target
supersymmetric signals
with large hadronic activity in conjunction with missing energy. 
The requirement for a minimum of
seven hard jets implies that this search should be most sensitive to
$\tilde g\tilde g$ and $\tilde g\tilde q$ production with subsequent decays into
top quarks. 
In this subsection we discuss how the exclusion limits shown in
the $(m_{\tilde u}, m_{\tilde g})$ plane depend on the two flavor-violating
parameters: the squark mass splitting $\Delta m$, and the squark mixing angle
$\theta^R_{23}$.
We take as benchmark
the case of a light stop and a decoupled scharm that are not mixed, 
with
$\Delta m = -500~\rm{GeV}$ and $\sin\theta^R_{23} = 0$.
In the upper panel of \figref{results_ATLASmultijet_del_m}
we show the reach of this ATLAS search for this scenario. 
In the other sub-figures of
\figref{results_ATLASmultijet_del_m}, we collect a representative set of results 
with various $\Delta m$ and $\theta^R_{23}$ values, which depict the interesting 
changes in the analysis reach due to flavor effects.

For cases in which the lightest squark is stop-like (scenarios of class
{\bf S.II} with $\Delta m < 0 $),
it is evident that the sensitivity of this search to gluino masses of about
$1.4\,{\rm TeV}$ is reduced by a non-zero stop-scharm mixing.
Due to flavor mixing, final states with charm quarks instead of top quarks
can arise, changing in this way the possible event topology and depleting some
of the multijet signal regions that require a large jet multiplicity.
A similar effect happens to the constraint on the mass of the lightest squark of
about $500\,{\rm GeV}$ when the gluino is heavy, with a mass ranging up to
$2\,{\rm TeV}$. This constraint originates mostly from direct squark-gluino
associated production, and the global jet multiplicity of the event is again
reduced when decays into charm quarks are allowed.

\begin{figure}
\centering
\includegraphics[width=0.45\textwidth]{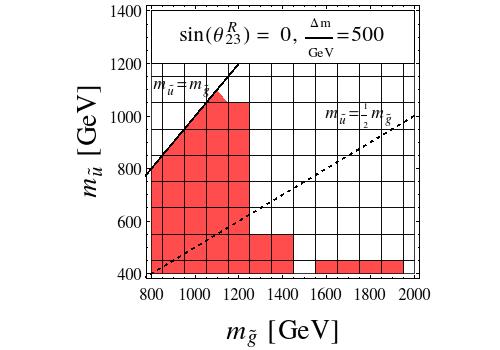}\\[1.5cm]
\includegraphics[width=0.45\textwidth]{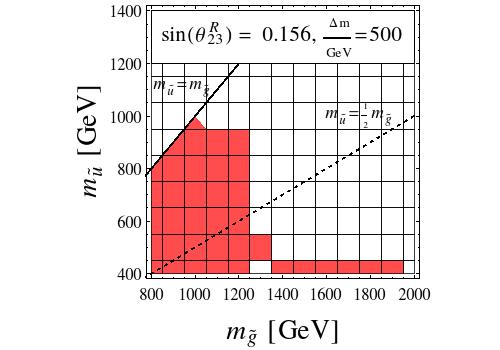}
\includegraphics[width=0.45\textwidth]{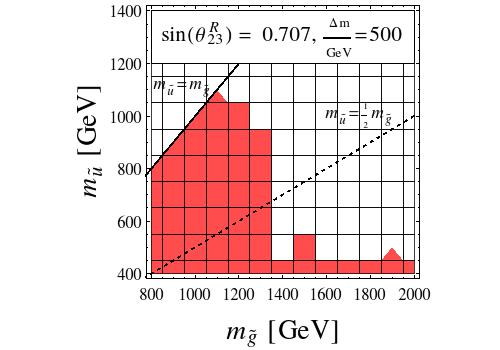}\\[1.5cm]
\includegraphics[width=0.45\textwidth]{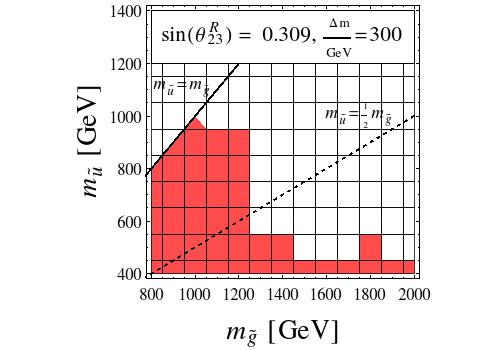}
\includegraphics[width=0.45\textwidth]{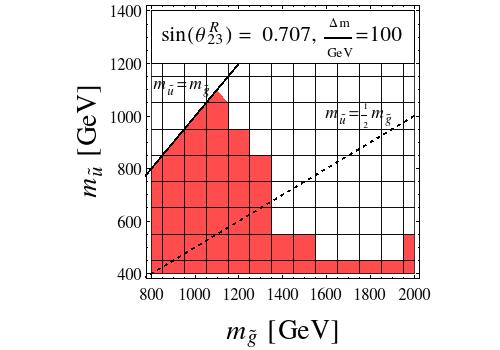}
\caption{
  Sensitivity of the ATLAS multijet + $\slashed{E}_T$ + lepton veto search
  of Ref.~\cite{Aad:2013wta} for different values of the $\Delta m$ and
  $\theta^R_{23}$ parameters (the exact values being indicated in the top bar of
  each subfigure). The excluded regions are shown in red in the $(m_{\tilde u},
  m_{\tilde g})$ plane, where $m_{\tilde u}$ is the mass of the lightest squark
  (a scharm-like squark here since $\Delta m > 0$).
  The upper panel describes the reference scenario in which
  the lightest squark is a pure scharm state and the stop is almost decoupled.
}
\label{fig:results_ATLASmultijet_del_p}
\end{figure}

Interestingly, close to the degeneracy line $m_{\tilde q} = m_{\tilde g}$, the
opposite effect occurs and the sensitivity increases when the mixing is turned
on. For $\sin\theta^R_{23}=0$, the two-body decay $\tilde g \to \tilde q t$ is
kinematically forbidden when $m_{\tilde g}-m_{\tilde t}<m_t$ so that the
dominant gluino decay mode proceeds through a three-body channel, $\tilde g\to
t\bar t\tilde \chi_1^0$. 
When the stop and the scharm mix, the two-body decay
mode  $\tilde g \to \tilde q c$ is open and dominates, even for relatively small
mixing angles. In this case, the top quarks from the subsequent squark
decay are more energetic than in the three-body decay case.
The decay products of these tops are therefore experimentally easier 
to detect so that the search sensitivity is enhanced.

In \figref{results_ATLASmultijet_del_p}, we focus on scenarios of type
{\bf S.I} where the lightest squark is scharm-like and $\Delta m > 0 $.
In the upper panel of the figure, we consider a reference scenario in which the
stop is decoupled and both squarks do not mix. In comparison with the
{\bf S.II} benchmark scenario, the search has a more limited reach. The reason
is twofold. On the one hand, the lighter scharm-like state is produced more
copiously than the the stop-like one via $\tilde g\tilde q$ production,
and on the other hand, the gluino branching fraction to charm-jets is
increased, leading to a depletion of the signal regions with a large jet
multiplicity. A non-vanishing stop-scharm mixing allows the gluino and the
scharm-like state to decay into top quarks, thus increasing the number of jets
in the event and repopulating the search signal regions by signal migration.
The impact of such a change is, however, rather modest, 
and becomes significant only for
large mixings.
This result is due to the small branching fractions of the gluino and
the scharm-like state into top quarks which are suppressed by the top mass, and
could therefore become substantial only for large values of the mixing angle.

\subsection{\boldmath CMS: single lepton + at least four jets (including at least
one $b$-jet) + $\slashed{E}_T$}

\begin{figure}
\centering
\includegraphics[width=0.45\textwidth]{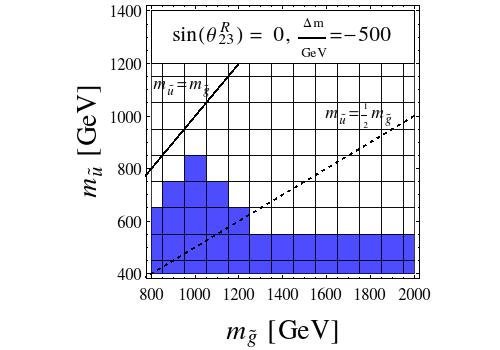}\\[1.5cm]
\includegraphics[width=0.45\textwidth]{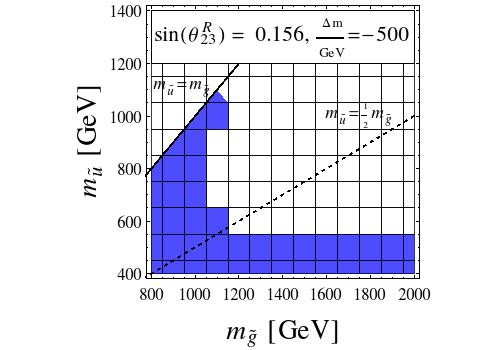}
\includegraphics[width=0.45\textwidth]{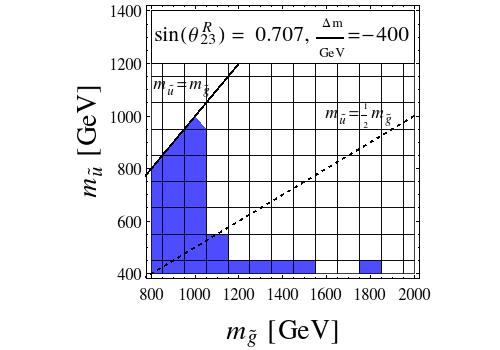}\\[1.5cm]
\includegraphics[width=0.45\textwidth]{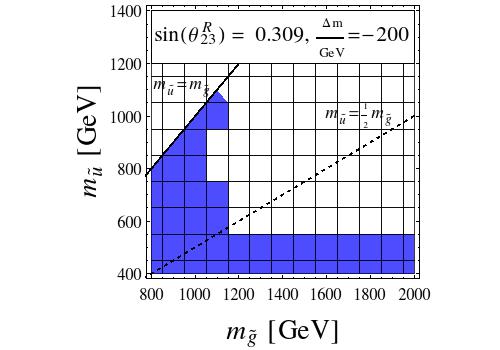}
\includegraphics[width=0.45\textwidth]{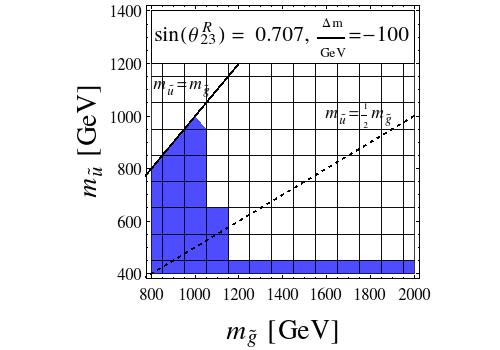}
\caption{
  Sensitivity of the CMS stop search in the single lepton mode
  of Ref.~\cite{Chatrchyan:2013xna} for different values of the $\Delta m$ and
  $\theta^R_{23}$ parameters (the exact values being indicated in the top bar of
  each subfigure). The excluded regions are shown in blue in the $(m_{\tilde u},
  m_{\tilde g})$ plane, where $m_{\tilde u}$ is the mass of the lightest squark
  (a stop-like squark here since $\Delta m < 0$).
  The upper panel describes the reference scenario in which
  the lightest squark is a pure stop state and the scharm is almost decoupled.
}
\label{fig:results_CMS1lep_del_m}
\end{figure}

\begin{figure}
\centering
\includegraphics[width=0.45\textwidth]{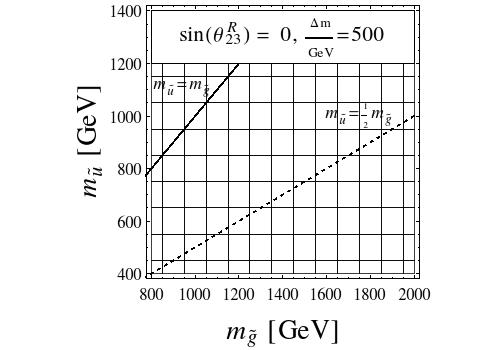}\\[1.5cm]
\includegraphics[width=0.45\textwidth]{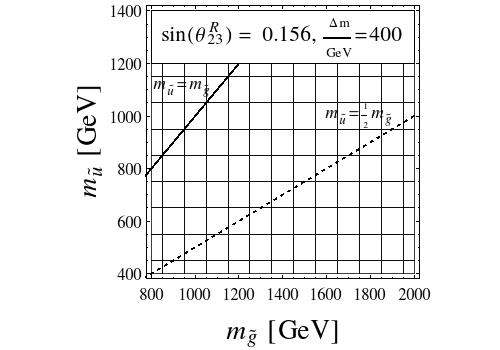}
\includegraphics[width=0.45\textwidth]{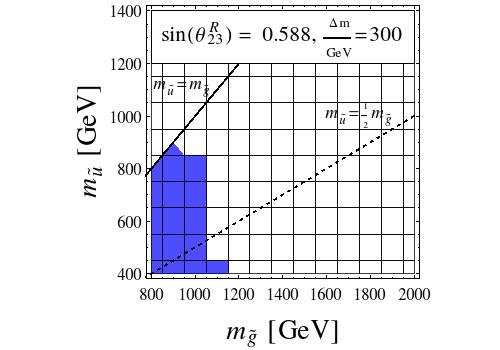}\\[1.5cm]
\includegraphics[width=0.45\textwidth]{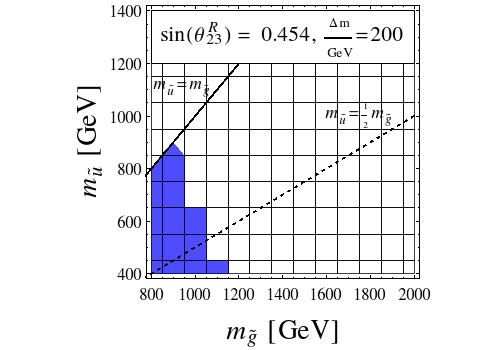}
\includegraphics[width=0.45\textwidth]{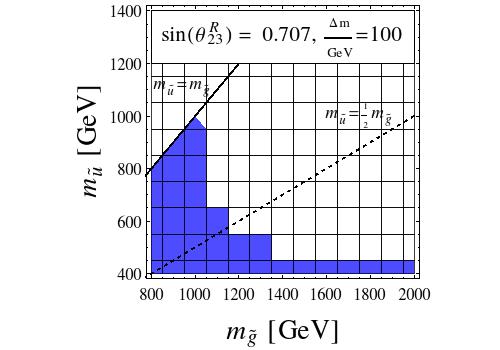}
\caption{
  Sensitivity of the CMS stop search in the single lepton mode
  of Ref.~\cite{Chatrchyan:2013xna} for different values of the $\Delta m$ and
  $\theta^R_{23}$ parameters (the exact values being indicated in the top bar of
  each subfigure). The excluded regions are shown in blue in the $(m_{\tilde u},
  m_{\tilde g})$ plane, where $m_{\tilde u}$ is the mass of the lightest squark
  (a scharm-like squark here since $\Delta m > 0$).
  The upper panel describes the reference scenario in which
  the lightest squark is a pure scharm state and the stop is almost decoupled.
}
\label{fig:results_CMS1lep_del_p}
\end{figure}

The CMS-SUS-13-011 search of Ref.~\cite{Chatrchyan:2013xna} has been designed to
look for the semileptonic decay of a stop pair. 
Consequently, we expect
this search to be rather insensitive to spectra where the lightest squark is
scharm-like (scenarios of type {\bf S.I}).
The results in \figref{results_CMS1lep_del_m}
and \figref{results_CMS1lep_del_p} show that this is indeed the case
with the exception of scenarios in which $0<\Delta m<200\,{\rm GeV}$ and/or in
which the mixing angle $\theta_{23}^R$ is large. 
In these cases, constraints on
the gluino masses of about 1~TeV find their origin in gluino-squark associated
production followed by their decay to two neutralinos and three quarks, 
at least two of which are tops.
In all the other cases with a light scharm-like squark,
the top mass severely suppresses any branching fraction to a final state
containing a top quark and therefore reduces the sensitivity of the search.

For a stop-like lightest squark (scenarios of class {\bf S.II}), the reach of
this search is much more significant. 
In addition to gluino pair-production,
direct squark pair-production can also yield a significant number of top-pairs.
As a result, light squark masses around 500\,GeV are
excluded independently of the gluino mass, unless the mixing angle is very
large. 
In the latter case (right column of \figref{results_CMS1lep_del_m}),
the signal regions are depleted when the squarks can decay not only to top quarks, 
but also to charm-jets which have the advantage of a larger phase-space.
Similarly to the ATLAS multijet search, the sensitivity to nearly degenerate
squark and gluino states increases with a non-zero stop-scharm mixing angle.

\subsection{\boldmath CMS: at least 3~jets + $\slashed{E}_T$ + lepton veto}

\begin{figure}
\centering
\includegraphics[width=0.45\textwidth]{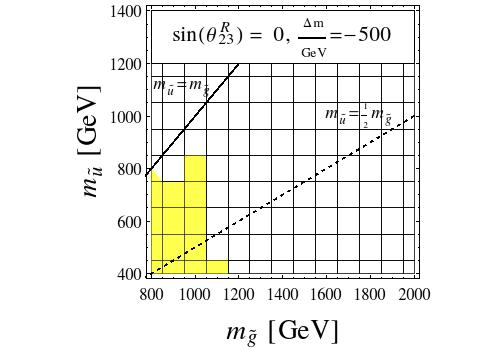}\\[1.5cm]
\includegraphics[width=0.45\textwidth]{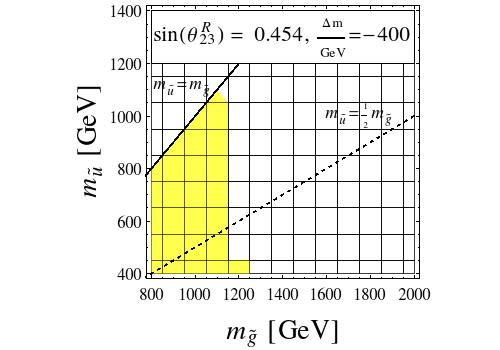}
\includegraphics[width=0.45\textwidth]{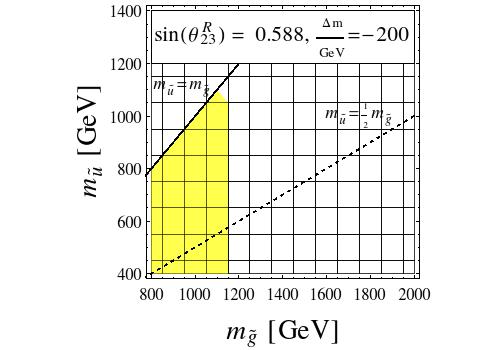}\\[1.5cm]
\includegraphics[width=0.45\textwidth]{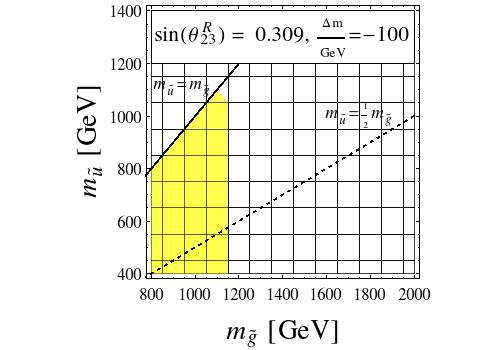}
\includegraphics[width=0.45\textwidth]{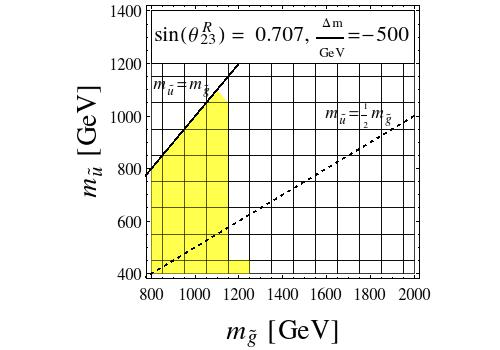}
\caption{
  Sensitivity of the CMS supersymmetry search in the multijet plus missing
  energy channel
  of Ref.~\cite{Chatrchyan:2014lfa} for different values of the $\Delta m$ and
  $\theta^R_{23}$ parameters (the exact values being indicated in the top bar of
  each subfigure). The excluded regions are shown in yellow in the $(m_{\tilde u},
  m_{\tilde g})$ plane, where $m_{\tilde u}$ is the mass of the lightest squark
  (a stop-like squark here since $\Delta m < 0$).
  The upper panel describes the reference scenario in which
  the lightest squark is a pure stop state and the scharm is almost decoupled.
}
\label{fig:results_CMS0lep_del_m}
\end{figure}

\begin{figure}
\centering
\includegraphics[width=0.45\textwidth]{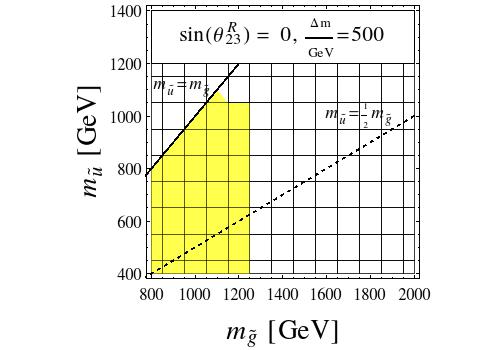}\\[1.5cm]
\includegraphics[width=0.45\textwidth]{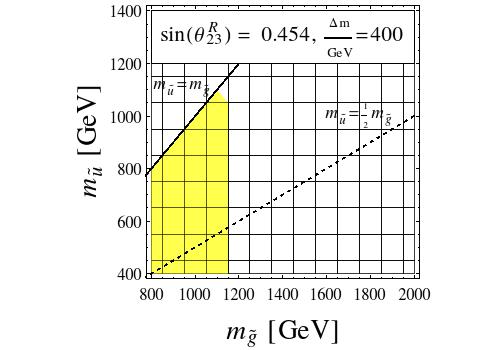}
\includegraphics[width=0.45\textwidth]{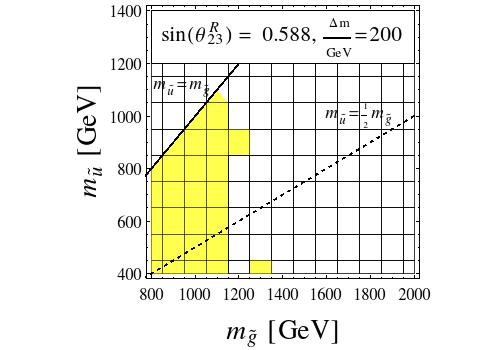}\\[1.5cm]
\includegraphics[width=0.45\textwidth]{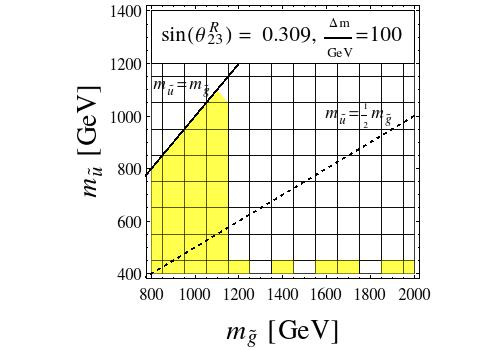}
\includegraphics[width=0.45\textwidth]{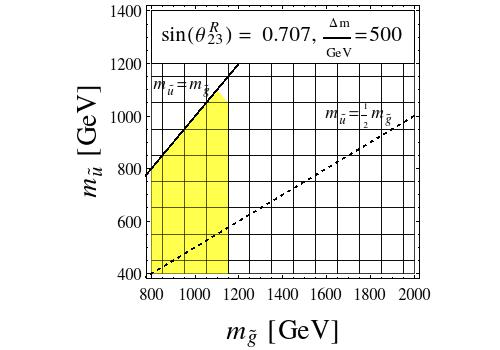}
\caption{
  Sensitivity of the CMS supersymmetry search in the multijet plus missing
  energy channel
  of Ref.~\cite{Chatrchyan:2014lfa} for different values of the $\Delta m$ and
  $\theta^R_{23}$ parameters (the exact values being indicated in the top bar of
  each subfigure). The excluded regions are shown in yellow in the $(m_{\tilde u},
  m_{\tilde g})$ plane, where $m_{\tilde u}$ is the mass of the lightest squark
  (a scharm-like squark here since $\Delta m > 0$).
  The upper panel describes the reference scenario in which
  the lightest squark is a pure scharm state and the stop is almost decoupled.
}
\label{fig:results_CMS0lep_del_p}
\end{figure}

Next, we consider the CMS-SUS-13-012 search of Ref.~\cite{Chatrchyan:2014lfa}
that is complementary to the previous CMS search. 
This search targets the purely hadronic decays 
of pair-produced superpartners, and therefore vetoes final-states which
include leptons.
The dependence of the sensitivity on the squark mass splitting and
the mixing angle is depicted in \figref{results_CMS0lep_del_m} and
\figref{results_CMS0lep_del_p}. 
Taking as a reference the non-mixing scenario of
class {\bf S.II} with a decoupled charm-squark (upper panel of
\figref{results_CMS0lep_del_m}), 
we observe that the sensitivity is enhanced in
all the other cases. This is a direct consequence of the relatively small
jet-multiplicity requirement of this CMS analysis alongside the lepton veto.
Those imply that events featuring charm-jets which come from
$\tilde g\tilde g$ and $\tilde g\tilde q$ production 
(whose respective cross-sections are large and mildly depend on the squark mass) 
are more likely to pass all the selection steps than those featuring top quarks.
The latter have indeed a non-negligible branching fraction into leptons,
and the top mass tends to limit the amount of missing energy in the events
(the neutralino $p_T$). 
Consequently, one expects that for the case in which the
lightest squark is stop-like, a non zero mixing will increase the reach of the
search, which is indeed the case as seen through \figref{results_CMS0lep_del_m}.

In an analogous way, one can explain the pattern of exclusion limits for scenarios
of class {\bf S.I} with a scharm-like lightest squark illustrated in
\figref{results_CMS0lep_del_p}. 
The gluino mass bound in the reference scenario is somewhat stronger here 
than in the stop-like case, but it is reduced for increasing mixing angles.

\subsection{ATLAS: scharm pair-production using charm-tagging + lepton veto}

\begin{figure}
\centering
\includegraphics[width=0.45\textwidth]{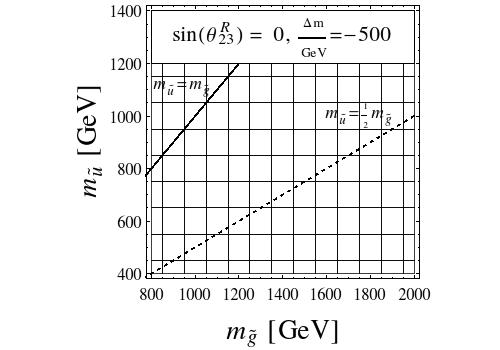}\\[1.5cm]
\includegraphics[width=0.45\textwidth]{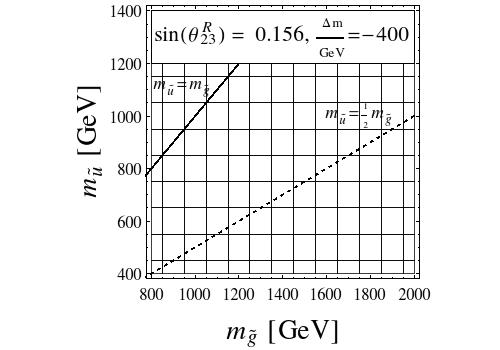}
\includegraphics[width=0.45\textwidth]{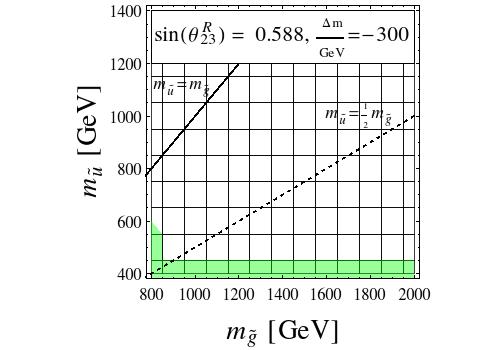}\\[1.5cm]
\includegraphics[width=0.45\textwidth]{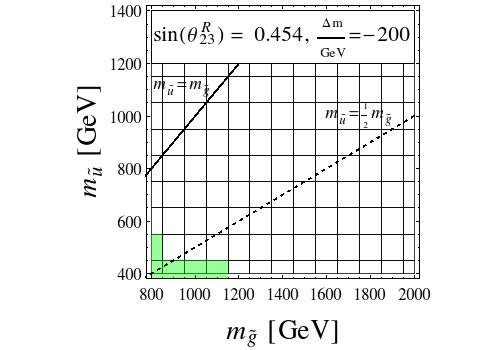}
\includegraphics[width=0.45\textwidth]{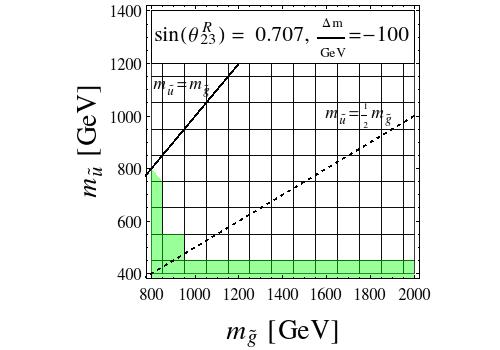}
\caption{
  Sensitivity of the ATLAS charm-squark search
  of Ref.~\cite{Aad:2015gna} for different values of the $\Delta m$ and
  $\theta^R_{23}$ parameters (the exact values being indicated in the top bar of
  each subfigure). The excluded regions are shown in green in the $(m_{\tilde u},
  m_{\tilde g})$ plane, where $m_{\tilde u}$ is the mass of the lightest squark
  (a stop-like squark here since $\Delta m < 0$).
  The upper panel describes the reference scenario in which
  the lightest squark is a pure stop state and the scharm is almost decoupled.
}
\label{fig:ATLAS_scharm_del_m}
\end{figure}

\begin{figure}
\centering
\includegraphics[width=0.45\textwidth]{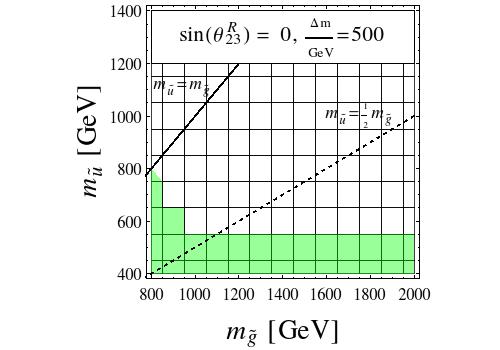}\\[1.5cm]
\includegraphics[width=0.45\textwidth]{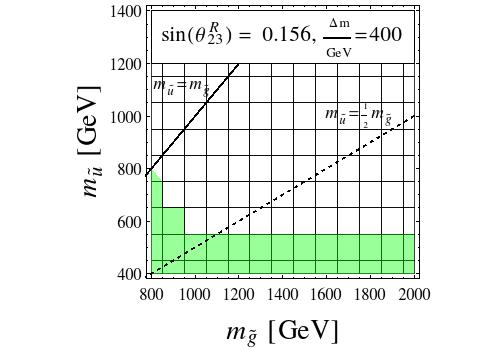}
\includegraphics[width=0.45\textwidth]{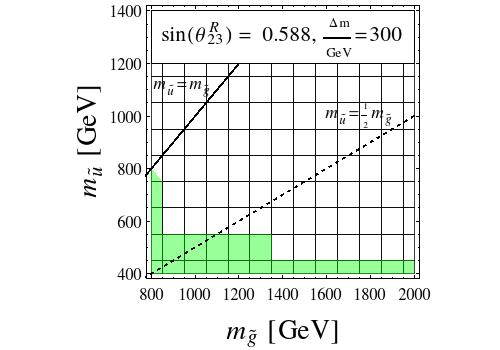}\\[1.5cm]
\includegraphics[width=0.45\textwidth]{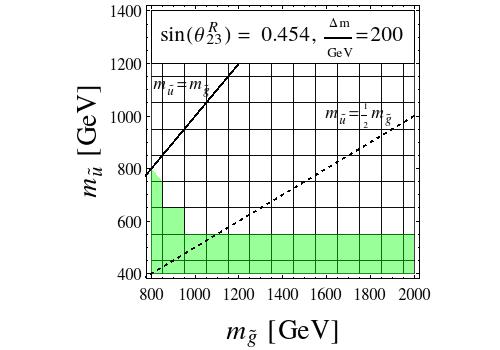}
\includegraphics[width=0.45\textwidth]{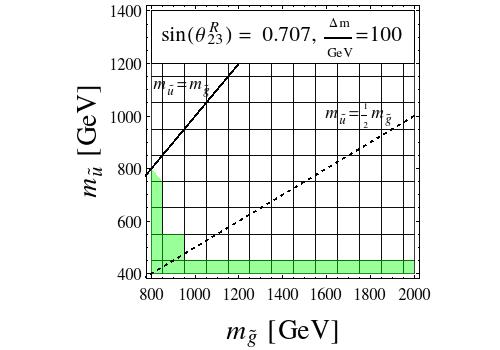}
\caption{
  Sensitivity of the ATLAS charm-squark search
  of Ref.~\cite{Aad:2015gna} for different values of the $\Delta m$ and
  $\theta^R_{23}$ parameters (the exact values being indicated in the top bar of
  each subfigure). The excluded regions are shown in green in the $(m_{\tilde u},
  m_{\tilde g})$ plane, where $m_{\tilde u}$ is the mass of the lightest squark
  (a scharm-like squark here since $\Delta m > 0$).
  The upper panel describes the reference scenario in which
  the lightest squark is a pure scharm state and the stop is almost decoupled.
}
\label{fig:ATLAS_scharm_del_p}
\end{figure}

The ATLAS-SUSY-2014-03 charm-squark search of Ref.~\cite{Aad:2015gna} targets
the production of a pair of charm squarks via a signature comprised of two
charm-tagged jets and missing transverse energy $\slashed E_T$. We recall that
our implementation of this search is very conservative and is only based on cross
sections and branching ratios. The sensitivity of this search to the model
studied in this work is presented in
\figref{ATLAS_scharm_del_m} and \figref{ATLAS_scharm_del_p}.

In case where the lightest squark is scharm-like (scenarios of class {\bf S.I}),
the search is sensitive to two specific regions of the parameter-space.
The first, with lower squark masses of about $500~\rm{GeV}$, is independent of
the gluino mass and extends up to  $m_{\tilde g} \sim 2~\rm{TeV}$.
In this domain, squark pair production is sufficient to yield an exclusion of
the model regardless of whether gluinos can be produced. In contrast, in the
second domain, the squarks are heavier, so that an exclusion of the model must
rely on the production of gluinos followed by their subsequent decays into
charm-jets. In this case, the reach depends on the gluino mass that limits the
production cross section. This feature also explains the effect of the different
mass splittings and mixings. For large enough positive mass splitting,
$\Delta m> 200~\rm{GeV}$, there is a light scharm-like state whose mixing
dependent branching fractions to charm-jets leads to an exclusion roughly
independent of the gluino mass. Although the exact value of the excluded mass
limit changes with the mixing angle, the general shape persists.
For smaller values of the squark mass splitting, both squark states become accessible
at the LHC, and with sufficiently large mixing, they contribute significantly to
the production of events with several charm-jets.

{The search loses sensitivity} if the lightest squark is a pure stop state
(scenarios of class {\bf S.II}). Consequently, the exclusion reach for negative
mass splittings strongly relies on the value of the mixing angle.
A large mixing indeed opens the possible supersymmetric decays into 
charm-jets. Alternatively, lowering the mass splitting also improves the reach
as it makes the heavier state, that is scharm-like, accessible.

\subsection{Combined Reach}
\begin{figure}
\centering
\includegraphics[width=0.45\textwidth]{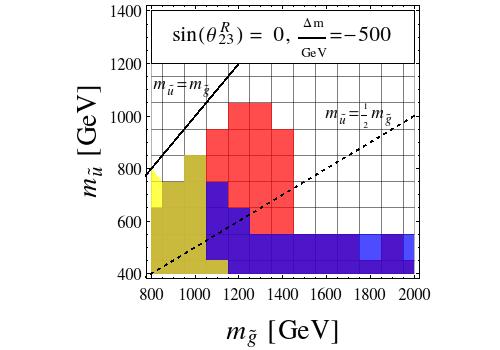}\\
\includegraphics[width=0.45\textwidth]{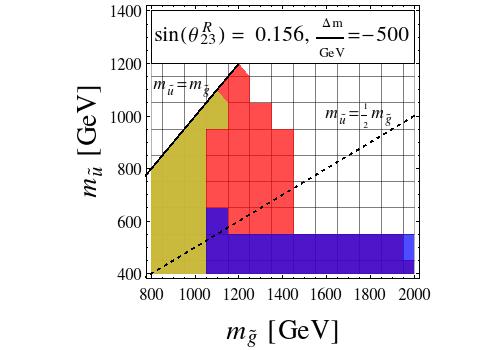}
\includegraphics[width=0.45\textwidth]{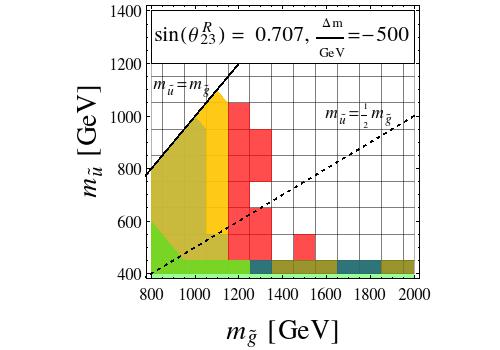}\\
\includegraphics[width=0.45\textwidth]{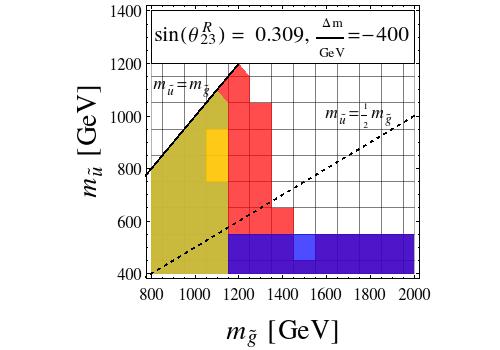}
\includegraphics[width=0.45\textwidth]{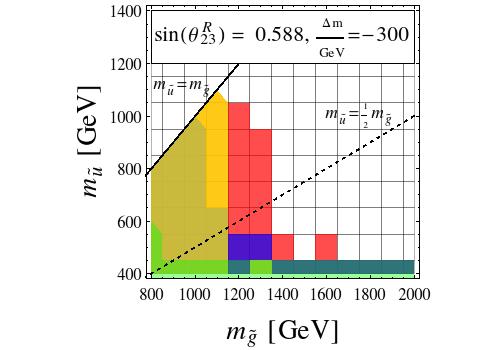}\\
\includegraphics[width=0.45\textwidth]{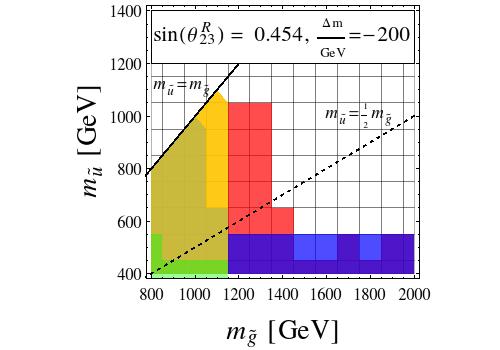}
\includegraphics[width=0.45\textwidth]{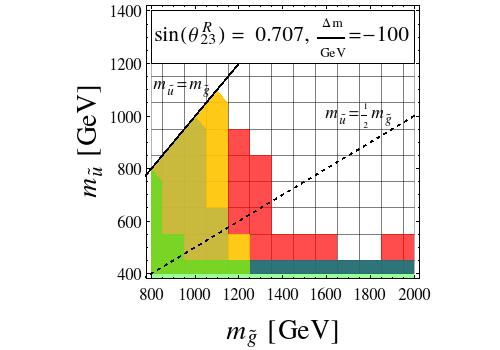}
\caption{
  Sensitivity of all four searches pursued in this work
  for different values of the $\Delta m$ and
  $\theta^R_{23}$ parameters (the exact values being indicated in the top bar of
  each subfigure). The excluded regions related to the ATLAS-SUSY-2013-04,
  CMS-SUS-13-011, CMS-SUS-13-012 and ATLAS-SUSY-2014-03 are respectively
  shown in red, blue, yellow and green in the $(m_{\tilde u},
  m_{\tilde g})$ plane, where $m_{\tilde u}$ is the mass of the lightest squark
  (a stop-like squark here since $\Delta m < 0$).
  The upper panel describes the reference scenario in which
  the lightest squark is a pure stop state and the scharm is almost decoupled.
}
\label{fig:combined_stop}
\end{figure}

\begin{figure}
\centering
\includegraphics[width=0.45\textwidth]{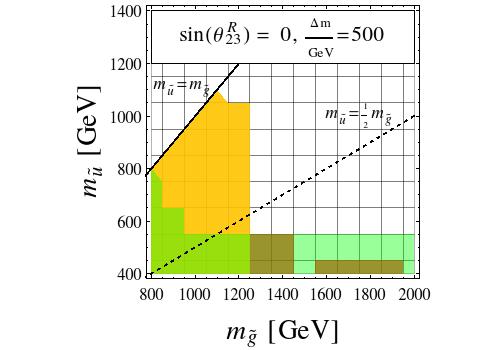}\\
\includegraphics[width=0.45\textwidth]{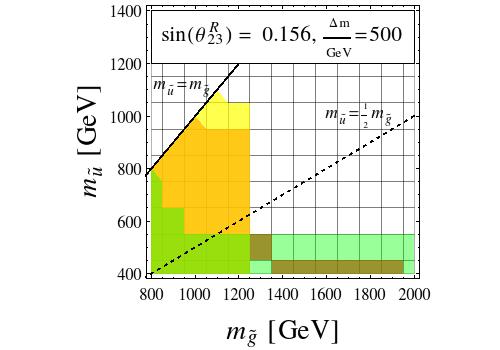}
\includegraphics[width=0.45\textwidth]{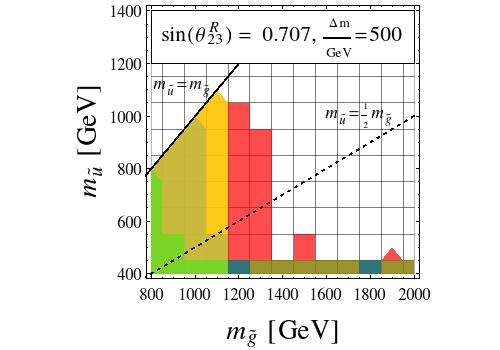}\\
\includegraphics[width=0.45\textwidth]{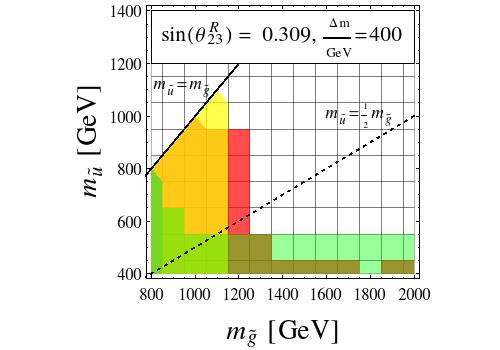}
\includegraphics[width=0.45\textwidth]{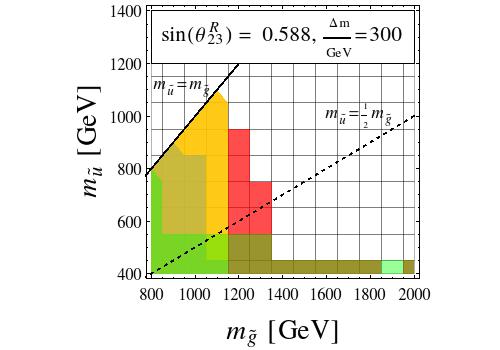}\\
\includegraphics[width=0.45\textwidth]{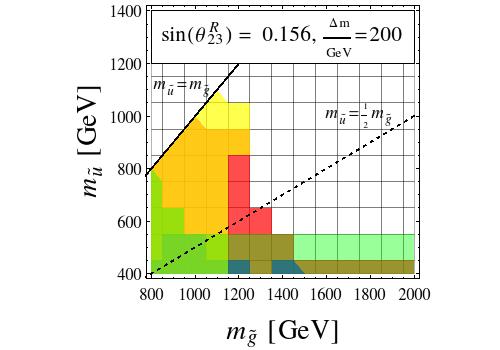}
\includegraphics[width=0.45\textwidth]{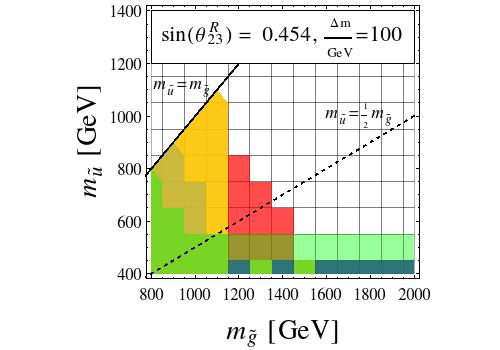}
\caption{
  Sensitivity of all four searches pursued in this work
  for different values of the $\Delta m$ and
  $\theta^R_{23}$ parameters (the exact values being indicated in the top bar of
  each subfigure). The excluded regions related to the ATLAS-SUSY-2013-04,
  CMS-SUS-13-011, CMS-SUS-13-012 and ATLAS-SUSY-2014-03 are respectively
  shown in red, blue, yellow and green in the $(m_{\tilde u},
  m_{\tilde g})$ plane, where $m_{\tilde u}$ is the mass of the lightest squark
  (a scharm-like squark here since $\Delta m > 0$).
  The upper panel describes the reference scenario in which
  the lightest squark is a pure scharm state and the stop is almost decoupled.
}
\label{fig:combined_scharm}
\end{figure}

It is interesting to discuss the combined reach of the four previously 
pursued searches for a few benchmarks points. To this end, we overlay all
four exclusion contours on top of each other, keeping the same color-coding as
for the individual results. We present the results in
\figref{combined_stop} and \figref{combined_scharm}.

Neglecting the effect of the LSP mass (taken here to be at its preferred value
for increasing the missing energy), the main result of the combined
reach figures is that the CMS search for three or more jets plus missing
energy and zero leptons, CMS-SUS-13-012~\cite{Chatrchyan:2014lfa}, leads to a
robust lower bound on the gluino mass of about 1.2~TeV regardless of the amount
of flavor mixing. Where the flavor mixing becomes significant, for
$\theta^R_{23} \gtrsim \pi/6$, an improvement in the reach is implied by
the ATLAS search in the multijet plus missing energy final state,
ATLAS-SUSY-2013-04~\cite{Aad:2013wta}, which leads to even stronger constraints
on the squark and gluino masses. Another noteworthy point is that the bound on
the lightest squark mass is always greater than about $400~\rm{GeV}$, reaching
even roughly $600~\rm{GeV}$ when this squark is primarily a scharm.
Finally, the CMS stop search in the single lepton final state,
CMS-SUS-13-011~\cite{Chatrchyan:2013xna}, is only relevant for stop-like
lightest squark while for the inverse scenario with a scharm-like lightest
squark, the same reach in the parameter space is this time covered roughly by
the ATLAS scharm-pair search, ATLAS-SUSY-2014-03~\cite{Aad:2015gna}.

\section{Conclusions}\label{sec:conclusions}
In this work we have studied the LHC constraints on the gluino and squark masses
in the presence of squark flavor violation.
The violation of the flavor symmetry has manifested itself in two interesting
and distinct manners. First, we have not assumed squark degeneracy,
and in particular, we have considered the case where the squarks of the first
and second generations have different masses.
Effectively, the first generation squarks, which are subject to the strongest
LHC constraints have been taken decoupled, while the would-be (right-handed)
scharm eigenstate has been allowed to be significantly lighter.
Second, we have allowed the squarks of the second and third generations,
in particular the (right-handed) scharm and stop, to mix.
Such a scenario has been studied in the past, however,
either under the assumption that the gluino is decoupled
and/or when the gluino flavor violating couplings are vanishing.
In the absence of non-MSSM structures, the former assumption is unnatural, 
and the latter is inconsistent with the basic flavor structure described above.

We can distinguish between two different sources of tuning. 
The first type is associated with the contribution of the would-be 
stop flavor eigenstate to the Higgs-boson mass.
In a previous work, this has been studied in models featuring a decoupled
gluino~\cite{Blanke:2013uia}, with the conclusion that the mixing could
slightly improve the level of required tuning. On this front, we have nothing
qualitatively new to add beyond the fact that the bounds on the stop- and scharm-like states have been slightly improved by the LHC experiments. 
The second type of tuning is related to the contributions of the gluino to 
the squark masses, where naturalness requires the gluino mass to be smaller than about twice of that of the squarks.  
This is particularly relevant to the above study due to the fact that the second generation squark masses are less constrained by the LHC searches.
Thus, the main purpose of our study has been to examine what are the bounds on the gluino-squark system within the above framework.

As the combined reach of the four LHC analyses that we have considered has
shown,
the fine-tuning requirement still allows for sizable stop-scharm mixing.
Equivalently, we have found that for various values of the mixing, there is a
wide range of unexcluded and relatively light
squark and gluino masses which satisfy the gluino-mass naturalness criterion.
In fact, for some mixings and mass splitting cases, the `natural' region in the squark-gluino mass plane could even become larger.

In the foreseen future, the experiments are expected to publish new results 
with an improved reach.
An improvement in the sensitivity to the above framework is obviously expected due to the increase in the center of mass energy. Furthermore, as the ATLAS collaboration has now effectively installed a new inner layer of pixel detector (IBL), its charm-tagging capabilities are expected to be upgraded, resulting in more efficient ways to look for charm squarks.
Moreover, we note that during the finalizing of this work, CMS has released several
analyses which target stop pair production and which are likely 
sensitive to gluinos as well.
The search with highest reach is presented in Ref.~\cite{CMS:2015kza}, with stop masses
excluded up to roughly~$750~\rm{GeV}$ (for an LSP defined as in our model).
Such an improved reach would cover additional parameter-space in 
our gluino-squark mass plane, however, it is expected to exhibit similar
characteristics once the mixing and mass-splittings are turned on.
As a result, the new CMS results stress the importance of flavor
even beyond the reach shown in this work.

Finally, we point out that we have not discussed here the impact of flavor violation
on low energy observables. 
We have also ignored the bounds coming from the Higgsino sector even though naturalness suggests that those should be at the bottom of the supersymmetric spectrum modulo, possibly, the dark matter candidate itself. These are beyond the scope of our study.

\acknowledgments
The authors are extremely grateful to David Cohen,
the administrator of the ATLAS-Technion grid project
for his extensive support. His personal support, and
assignment of resources have greatly facilitated in
the progress of this work.
We also thank Monica d'Onofrio and Tomasso Lari
for their help with the reimplementation of the ATLAS-SUSY-2013-04
search.
BF is supported by the Theory-LHC-France initiative of the
CNRS (INP/IN2P3). IG is supported by NSF grant PHY-1316792.
The work of GP is supported by BSF, ERC and ISF grants.

\appendix
\section{Implementation and Validation of ATLAS-SUSY-2013-04 
in {\sc MadAnalysis}~5}
\label{sec:validation}
The recast of the ATLAS-SUSY-2013-04 search makes use of the
{\sc MadAnalysis}~5 framework together with the {\sc MA5tune} version of
{\sc Delphes} described in Ref.~\cite{Dumont:2014tja}. 
Our implementation therefore
relies on the interfacing of {\sc Delphes} as implemented in the version 1.1.12 of
{\sc MadAnalysis}~5. The necessary
configuration file of {\sc Delphes} can be found on the public analysis
database webpage,\\
\verb+  http://madanalysis.irmp.ucl.ac.be/wiki/PublicAnalysisDatabase+ .\\
The validation of the analysis implementation has been achieved by relying on a
set of benchmark scenarios belonging to the gluino-stop (off-shell) model
introduced in the ATLAS publication~\cite{Aad:2013wta}. In this case, the SM is
supplemented by a gluino and a neutralino, and the gluino is enforced to decay
into a top-antitop final state via an off-shell top squark. For validation
purposes, we generate events following the procedure of the ATLAS collaboration,
using the {\sc Herwig++} program~\cite{Bahr:2008pv} for the simulation of the
hard process, the parton showering and the hadronization. The supersymmetric
spectrum file has been provided by the ATLAS collaboration {\it via}
{\sc HepData} and the {\sc Herwig++} configuration that we have used can be
obtained from the {\sc MadAnalysis}~5 webpage.

\renewcommand\arraystretch{1.15}
\begin{table}
  \renewcommand{\tabcolsep}{0.30cm}
  \begin{tabular}{l||c|c||c|c}
    \multirow{2}{*}{\textbf{Selection step}} & \multicolumn{2}{c||}{\textbf{Events counts}} &\multicolumn{2}{c}{\textbf{Relative change}}\\
    & {\sc MadAnalysis 5}& ATLAS & {\sc MadAnalysis 5} & ATLAS \\
  \hline
    Initial number of events   & 206.3& 206.3&         &         \\
    6 jets with $E_T > 45$~GeV & 157.0& 168  & -23.9\% & -18.6\% \\
    lepton veto                &  88.1&  78  & -43.9\% & -53.6\% \\
  \hline
    8 jets ($p_T>50$~GeV)          & 18.3 & 16.3 & -79.2\% & -79.1\% \\
    $\slashed{E}_T/\sqrt{H_T}>4$~GeV$^{1/2}$ & 15.4 & 14.1 & -15.8\% & -13.5\% \\
    $\to$ without $b$-tags         & 0.97 & 0.85 & -93.7\% & -94.0\% \\
    $\to$ with 1 $b$-tag           & 3.80 & 3    & -75.3\% & -78.7\% \\
    $\to$ with 2 $b$-tags          & 10.7 & 11   & -30.5\% & -22.0\% \\
  \hline
    9 jets ($p_T>50$~GeV)          & 11.6&  9.6  & -86.8\% & -87.7\% \\
    $\slashed{E}_T/\sqrt{H_T}>4$~GeV$^{1/2}$ & 9.32&  8.0  & -19.7\% & -16.7\% \\
    $\to$ without $b$-tags         & 0.70&  0.33 & -92.5\% & -95.9\% \\
    $\to$ with 1 $b$-tag           & 1.93&  1.7  & -79.3\% & -78.8\% \\
    $\to$ with 2 $b$-tags          & 6.68&  6.5  & -28.3\% & -18.8\% \\
  \hline
    $\geq10$ jets ($p_T>50$~GeV)   & 6.99& 5.7  & -92.1\% & -92.7\% \\
    $\slashed{E}_T/\sqrt{H_T}>4$~GeV$^{1/2}$ & 5.34& 4.7  & -23.6\% & -17.5\% \\
  \hline
  \hline
    7 jets ($p_T>80$~GeV)          & 9.72& 7.53 & -88.1\% & -87.7\% \\
    $\slashed{E}_T/\sqrt{H_T}>4$~GeV$^{1/2}$ & 8.07& 6.25 & -17.0\% & -17.0\% \\
    $\to$ without $b$-tags         & 0.47& 0.31 & -94.2\% & -95.0\% \\
    $\to$ with 1 $b$-tag           & 1.73& 1.3  & -78.6\% & -79.2\% \\
    $\to$ with 2 $b$-tags          & 5.86& 5.1  & -27.4\% & -18.4\% \\
  \hline
    $\geq8$ jets ($p_T>80$~GeV)    & 4.72& 3.2  & -94.6\% & -95.9\% \\
    $\slashed{E}_T/\sqrt{H_T}>4$~GeV$^{1/2}$ & 3.65& 2.6  & -22.7\% & -18.8\% \\
    $\to$ without $b$-tags         & 0.06& 0.13 & -98.4\% & -95.0\% \\
    $\to$ with 1 $b$-tag           & 0.72& 0.55 & -80.3\% & -78.8\% \\
    $\to$ with 2 $b$-tags          & 2.87& 2.1  & -21.4\% & -19.2\% \\
\end{tabular}
\caption{Summary of yields for a gluino-stop off-shell scenario in which the
gluino and neutralino masses have been fixed to 1100 and 400~GeV, respectively.
The results obtained with {\sc MadAnalysis}~5 are compared to the official ATLAS
results, both in terms of event counts and efficiencies computed from the
number of events before and after each of the selection steps.
\label{tab:atlasvalidation}}
\end{table}
\renewcommand\arraystretch{1.}

\begin{figure}
 \includegraphics[width=0.48\textwidth]{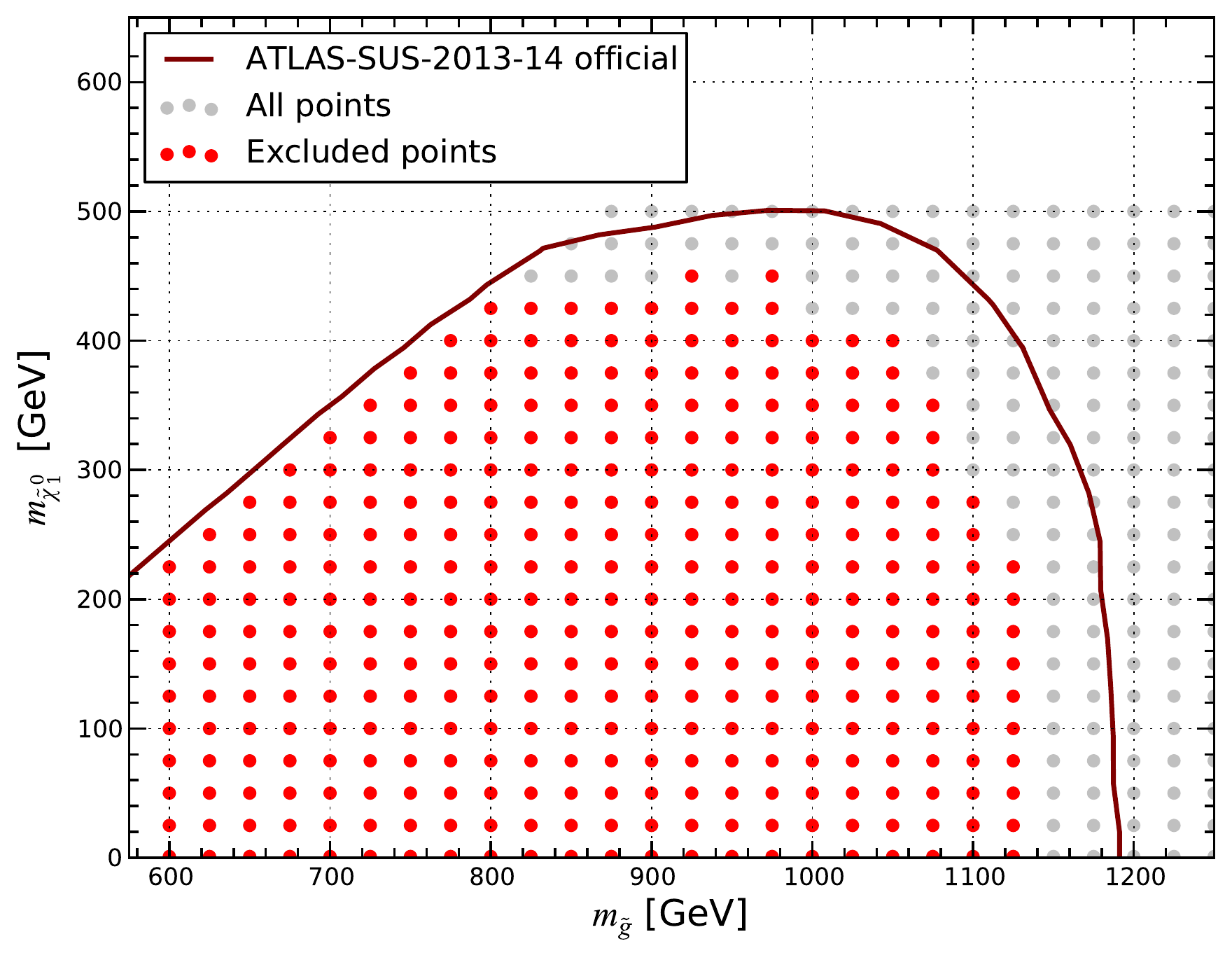}
 \caption{95\% CL exclusion limit for gluino-stop off-shell scenarios presented as
contours in the gluino-neutralino mass plane. We compare results obtained with
{\sc MadAnalysis}~5 (dots) to the official ATLAS results (solid line).
\label{fig:atlasvalidation}}
\end{figure}

In Table~\ref{tab:atlasvalidation}, we compare the ATLAS results for the cut-flow
counts to those obtained with our reimplementation of the ATLAS-SUSY-2013-04
analysis in {\sc MadAnalysis}~5. We present the surviving number of events after
each step of the selection strategy for the 13 signal regions under consideration
and for a scenario in which the gluino mass is set to
1100~GeV and the neutralino mass to 400~GeV. We have found that all
selection steps are properly described by the {\sc MadAnalysis}~5
implementation, the agreement reaching the level of about 10\%. In
Figure~\ref{fig:atlasvalidation}, we move away from the chosen benchmark
scenario and vary the gluino and neutralino masses freely, enforcing however that the gluino
decay channel into a top-antitop pair and a neutralino stays open. We observe
that our machinery allows us to reproduce the ATLAS official bounds (obtained
from {\sc HepData}) at the 50~GeV level, which is acceptable on the basis of
the limitations of our procedure mentioned in Section~\ref{sec:generalities}.
The {\sc MadAnalysis}~5 implementation of the analysis can be obtained from
{\sc Inspire}~\cite{inspire}.

\bibliography{GFV}
\end{document}